\documentclass[sigconf]{acmart}
\usepackage{multirow}
\usepackage{subcaption}
\usepackage{comment}

\usepackage[compact]{titlesec}

\titlespacing{\section}{0pt}{2ex}{1ex}
\titlespacing{\subsection}{0pt}{1ex}{0ex}
\titlespacing{\subsubsection}{0pt}{0.5ex}{0ex}

\usepackage{enumitem}
\setenumerate[1]{itemsep=0pt,partopsep=0pt,parsep=\parskip,topsep=3pt}
\setitemize[1]{itemsep=0pt,partopsep=0pt,parsep=\parskip,topsep=3pt}
\setdescription{itemsep=0pt,partopsep=0pt,parsep=\parskip,topsep=3pt}
\setlist{leftmargin=4.5mm}
\AtBeginDocument{%
  \providecommand\BibTeX{{%
    \normalfont B\kern-0.5em{\scshape i\kern-0.25em b}\kern-0.8em\TeX}}}

\copyrightyear{2020}
\acmYear{2020}
\setcopyright{acmcopyright}\acmConference[CIKM '20]{Proceedings of the 29th ACM
International Conference on Information and Knowledge Management}{October
19--23, 2020}{Virtual Event, Ireland}
\acmBooktitle{Proceedings of the 29th ACM International Conference on Information
and Knowledge Management (CIKM '20), October 19--23, 2020, Virtual Event, Ireland}
\acmPrice{15.00}
\acmDOI{10.1145/3340531.3412681}
\acmISBN{978-1-4503-6859-9/20/10}

\begin{document}
\fancyhead{}
\title{A Deep Prediction Network for Understanding\\Advertiser Intent and Satisfaction}


\author{Liyi Guo$^1$, Rui Lu$^2$, Haoqi Zhang$^1$, Junqi Jin$^2$, Zhenzhe Zheng*$^1$, Fan Wu$^1$, Jin Li$^2$, Haiyang Xu$^2$
\and
Han Li$^2$, Wenkai Lu$^3$, Jian Xu$^2$, Kun Gai$^2$}
\affiliation{
\institution{$^1$Shanghai Jiao Tong University, $^2$Alibaba Group, $^3$Tsinghua University}
\and
\{liyiguo1995, zhanghaoqi39, zhengzhenzhe\}@sjtu.edu.cn, fwu@cs.sjtu.edu.cn
\and
\{dashi.lr, junqi.jjq, echo.lj, shenzhou.xhy, lihan.lh, xiyu.xj\}@alibaba-inc.com
\and
lwkmf@mail.tsinghua.edu.cn, jingshi.gk@taobao.com}
\thanks{This work was supported in part by National Key R\&D Program of China No. 2019YFB2102200, in part by Alibaba Group through Alibaba Innovation Research Program, in part by China NSF grant No. 61902248, 61972252, 61972254, 61672348, and 61672353, in part by Joint Scientific Research Foundation of the State Education Ministry No. 6141A02033702, in part by the Open Project Program of the State Key Laboratory of Mathematical Engineering and Advanced Computing No. 2018A09. The opinions, findings, conclusions, and recommendations expressed in this paper are those of the authors and do not necessarily reflect the views of the funding agencies or the government.}
\thanks{*Z. Zheng is the corresponding author.}
\renewcommand{\shortauthors}{Guo et al.}

\begin{abstract}
For e-commerce platforms such as Taobao and Amazon, advertisers play an important role in the entire digital ecosystem: their behaviors explicitly influence users' browsing and shopping experience; more importantly, advertiser's expenditure on advertising constitutes a primary source of platform revenue. Therefore, providing better services for advertisers is essential for the long-term prosperity for e-commerce platforms. To achieve this goal, the ad platform needs to have an in-depth understanding of advertisers in terms of both their marketing intents and satisfaction over the advertising performance, based on which further optimization could be carried out to service the advertisers in the correct direction. In this paper, we propose a novel Deep Satisfaction Prediction Network (DSPN), which models advertiser intent and satisfaction simultaneously. It employs a two-stage network structure where advertiser intent vector and satisfaction are jointly learned by considering the features of advertiser's action information and advertising performance indicators. Experiments on an Alibaba advertisement dataset and online evaluations show that our proposed DSPN outperforms state-of-the-art baselines and has stable performance in terms of AUC in the online environment. Further analyses show that DSPN not only predicts advertisers' satisfaction accurately but also learns an explainable advertiser intent, revealing the opportunities to optimize the advertising performance further.
\end{abstract}

\ccsdesc[500]{Information systems~Online advertising}
\ccsdesc[500]{Applied computing~Electronic commerce}

\keywords{E-commerce; Display Advertisement; Advertiser Intent Identification; Advertiser Satisfaction Prediction}

\maketitle
\begin{small}
\textbf{ACM Reference Format:}\\
Liyi Guo, Rui Lu, Haoqi Zhang, Junqi Jin, Zhenzhe Zheng, Fan Wu, Jin
Li, Haiyang Xu, Han Li, Wenkai Lu, Jian Xu, and Kun Gai. 2020. A Deep
Prediction Network for Understanding Advertiser Intent and Satisfaction.
In \emph{Proceedings of the 29th ACM International Conference on Information
and Knowledge Management (CIKM '20), October 19–23, 2020, Virtual Event, Ireland}. ACM, New York, NY, USA, 8 pages. https://doi.org/10.1145/3340531.3412681
\end{small}

\section{Introduction}
\begin{figure}[!t]
 \centering
  \includegraphics[width=1.0\linewidth]{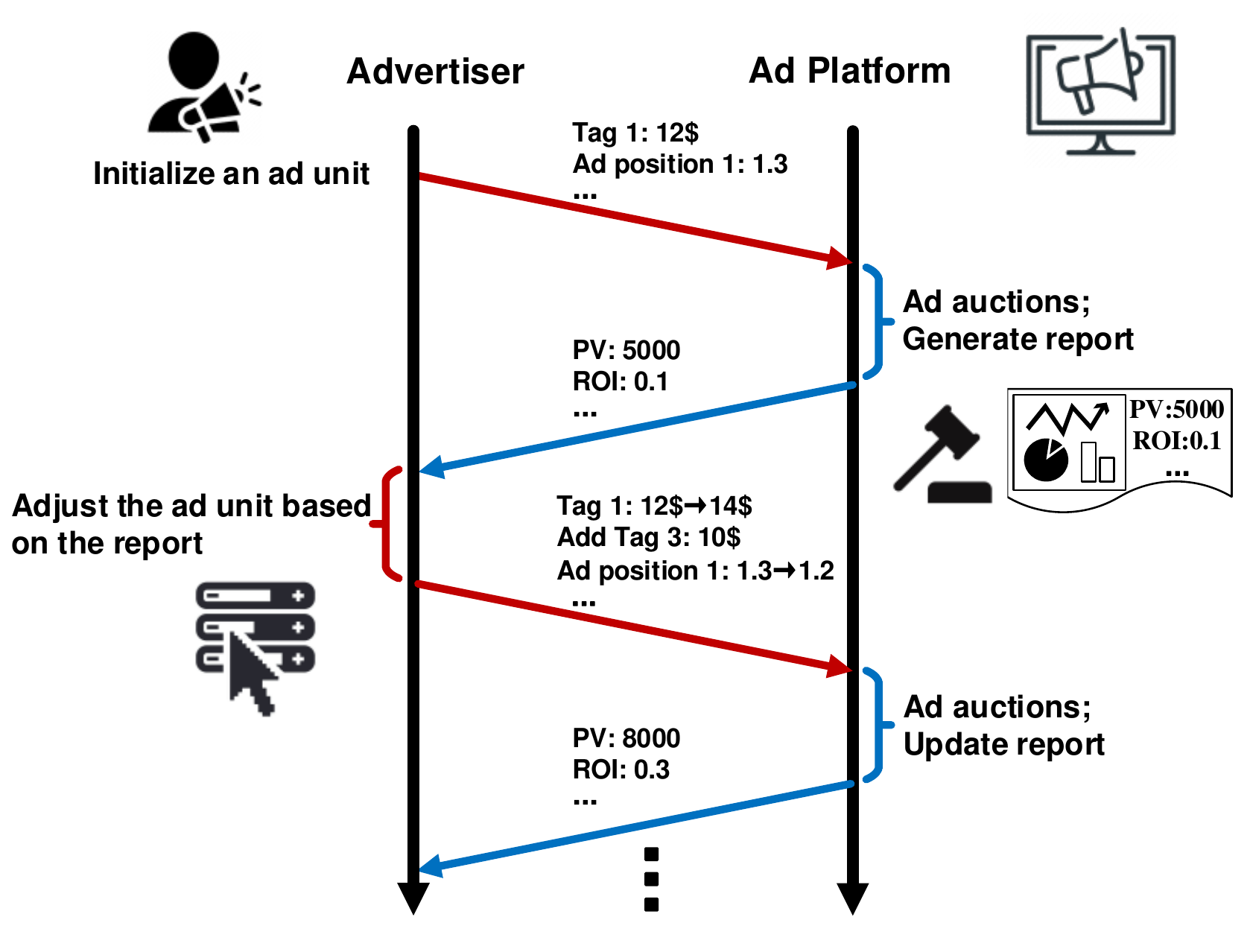}
  \caption{An representative interaction process between an advertiser and the ad platform.}
  \label{fig:system}
\end{figure}

As online-shopping growing rapidly in popularity, e-commerce platforms such as Taobao and Amazon have become a primary place for users to search and purchase products~\cite{evans2008economics, evans2009online, goldfarb2011online}. With a huge number of potential users in online platforms, an increasingly large number of sellers rely on e-commerce platforms for advertising their products. Although advertising expenses constitute a major revenue income of e-commerce platforms, little attention has been given to investigate advertiser's behavior, intent or satisfaction either from academia or industry community. Previous studies mainly focused on modeling and understanding the behaviors of users, such as widely-studied Click-Through Rate (CTR) prediction~\cite{covington2016youtube, cheng2016wide, gai2017learning, guo2017deepfm, zhou2018DIN, zhou2019DIEN} and user intent or satisfaction estimation in diverse contexts~\cite{chapelle2011intent, wang2014modeling, mehrotra2017deep, su2018user, guo2019buying, mehrotra2019jointly}. 
As an ad platform can be modeled as a two-sided market between users and advertisers~\cite{evans2009online}, ignoring the actions and feedback from advertisers would significantly reduce the efficiency of user-advertiser matching, and deteriorate the performance of ad systems in the long term. From a recently launched survey for advertisers in Taobao displaying ad platform, we observe that around half of the new advertisers would leave the ad platform after ten days, due to their marketing goals are not well satisfied. Therefore, in this work, we initialize a new research direction: focusing on advertiser intent understanding and advertiser satisfaction prediction, which would be leveraged to further improve the performance of ad systems.

Advertisers achieve their ad marketing goals through taking different actions for ad campaigns, resulting in frequent and rich interactions with the ad platforms. We illustrate one representative interaction process between an advertiser and the ad platform in Figure~\ref{fig:system}. The advertisers launch an ad campaign for a specific product by initializing an \emph{ad unit}: adding demographic tags of targeting users and setting basic bid prices. The ad unit participates in a series of ad auctions over a period of time, and the statistical indicators about the ad performance, such as CTR, CVR (Conversion Rate), ROI (Return of Investment), and etc, are recorded in a report. After observing the performance report, the advertiser would take several adjusting actions, such as adding/deleting some tags or modifying bid prices, to further improve the ad performance under the guideline of her marketing goal. The ad auction mechanisms would react to the adjustment of advertiser and generate new ad performance. We observe that if an advertiser is satisfied with the updated ad performance, in most cases she would like to continue to invest in advertising; otherwise, she would eventually close the ad campaign and leave the platform. Thus, the long-term revenue of ad platforms largely depends on whether the advertisers are satisfied with the performance of their ad campaigns. It is highly necessary to provide algorithms for the ad platforms to predict advertiser's satisfaction.

\begin{figure}[t]
\centering

    \begin{subfigure}{0.23\textwidth}
    \centering
        \includegraphics[width=1.0\linewidth]{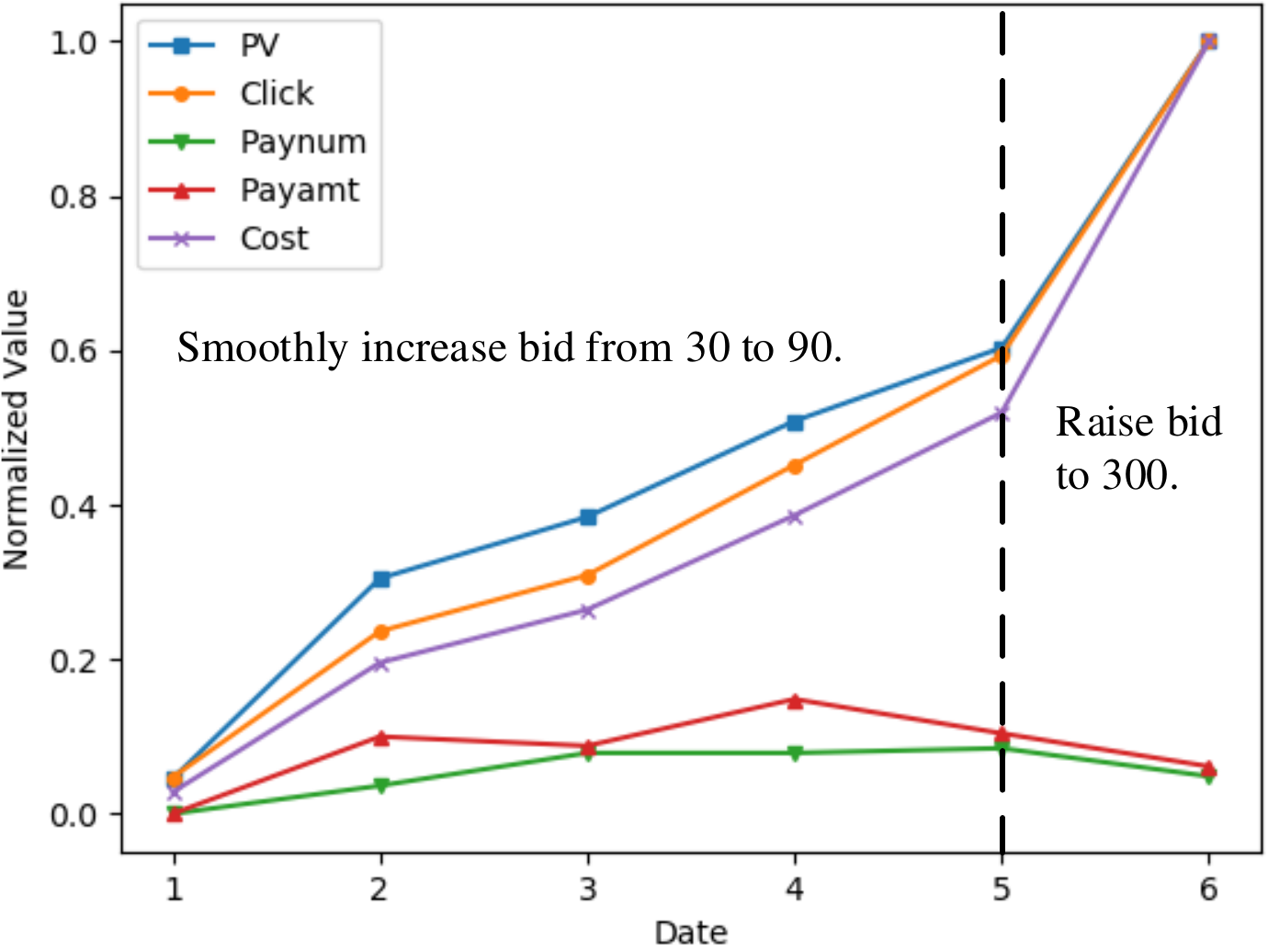}
        \caption{{Maximizing Impression.}}
        \label{fig:intro_example_1}
    \end{subfigure}
    \begin{subfigure}{0.23\textwidth}
    \centering
        \includegraphics[width=1.0\linewidth]{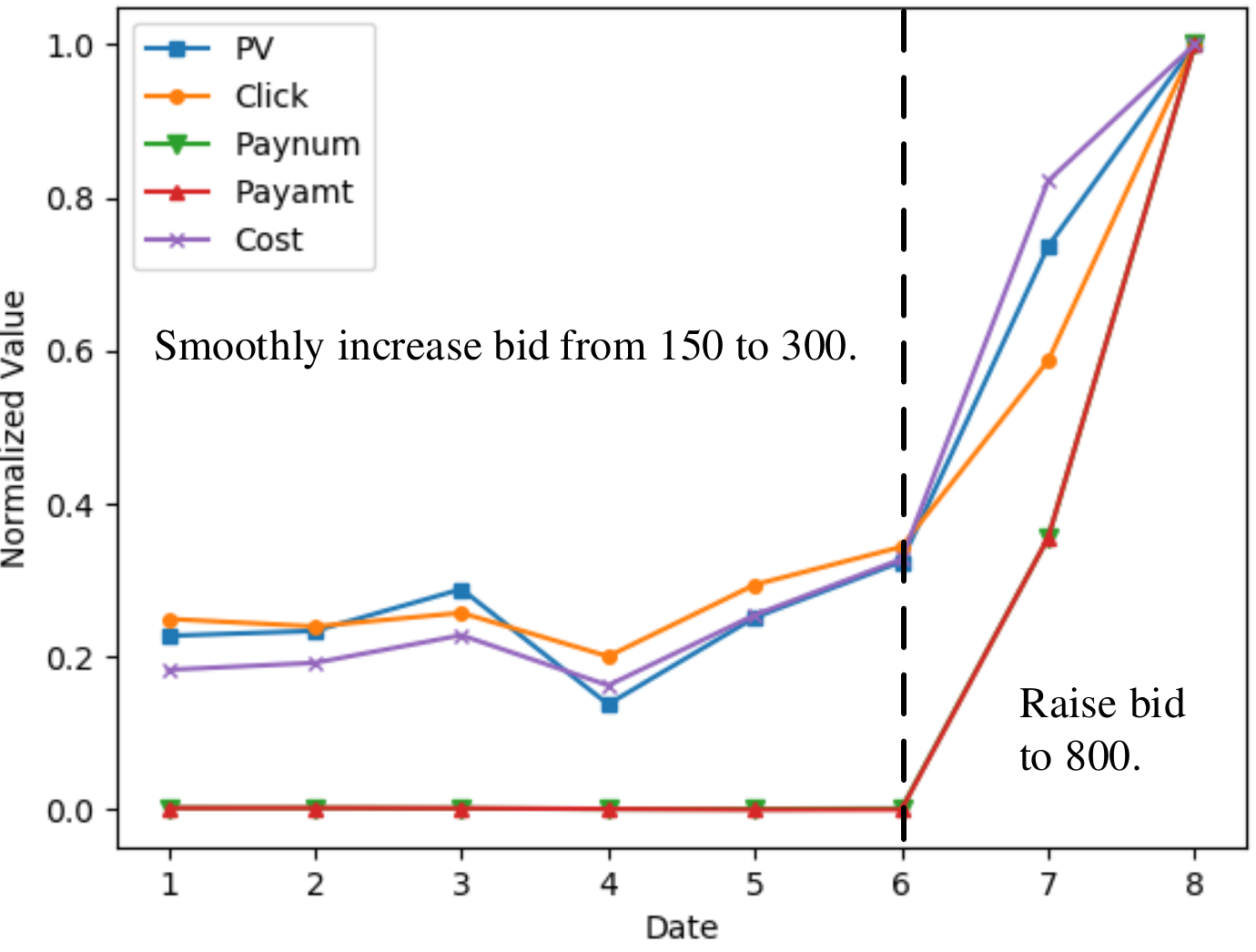}
        \caption{Maximizing Revenue.}
        \label{fig:intro_example_2}
    \end{subfigure}%
    
\caption{Advertisers realize their various intents through increasing bid prices. To understand advertisers, we should get more details to model and identify advertisers' intents.}
\label{fig:intent_example}
\end{figure}

During the advertising campaign, heterogeneous advertisers would have quite different preferences over the performance indicators, demonstrating various marketing intents. Although our case study is in a display advertising system with cost per click (CPC), advertisers still have different marketing intents beyond Click numbers. For example, for new launched products, advertisers tend to reach users as many as possible in a given time period, in which PV (Page View) and Click number often act as the key factors; for the advertisers with the goal of maximizing transaction revenue, they would pay more attention to indicators of Paynum (Pay Number) and Payamt (Pay Amount). Figure~\ref{fig:intent_example} illustrates the examples of these two advertisers realizing their different intents through a series of actions. The first advertiser would like to maximize user impressions, while the second advertiser attempts to maximize revenue. To provide better personalized ad services and optimize ad system performance in correct directions, the ad platform needs to have a deep understanding of the marketing intents of advertisers.

In this paper, we propose a novel advertiser-side Deep Satisfaction Prediction Network (DSPN), which jointly models the advertiser intent and predicts advertiser satisfaction. Specifically, DSPN is a supervised learning framework that employs a two-stage hierarchical structure: through transformation of features, attention mechanism and recurrent neural networks, the first stage leverages both advertiser's action information and various ad performance indicators to learn the potential intent, which is further fed into the second stage; the second stage models the relation between advertiser intent vector and satisfaction explicitly in a principled way. Based on the above architecture, DSPN automatically learns the connections among raw data, advertiser intent and metrics related to advertiser satisfaction in e-commerce. Our contributions of this work can be summarized as follows:

(1) To the best of our knowledge, we are the first to consider advertiser-side intent identification and satisfaction prediction. We formally define advertiser intent as a multi-dimension weight vector over ad performance indicators, which can be derived from a mathematical optimization problem. We evaluate the advertiser's satisfaction by a new metric based on the change of investment on ad campaigns.

(2) We point out the challenges in directly applying user-sided models to the advertiser-sided problems since they did not model the intent in a unified continuous space. Thus, we propose a Deep Satisfaction Prediction Network (DSPN) which jointly model advertiser intent and satisfaction. By employing action fusion layer and recurrent network structures, DSPN effectively extracts information critical for understanding advertiser intent and satisfaction, from diverse data features underlying the interaction between advertisers and the ad platform.

(3) We conduct extensive experiments on an Alibaba advertisement dataset. Results verify the effectiveness of DSPN both in satisfaction prediction and intent vector generation. The generated intent vector not only exhibits nice interpretation for advertiser's behavior but also reveals the potential optimization objectives to further improve the advertiser satisfaction and ad system performance. Moreover, DSPN has now been deployed in Alibaba online advertiser churn prediction system and serves daily business requirements.

\section{Related Work}
\subsection{User Intent Modeling and Satisfaction Prediction}
In traditional information retrieval literature, there are extensive existing work in predicting user satisfaction with a searching query under different user intents~\cite{mehrotra2017deep, chapelle2011intent, wang2014modeling, interpreting}. In \cite{interpreting}, the authors considered the keywords submitted in an ad auction as a direct expression of intent and use them to determine the relevance of the ad and user query. Recently, researchers initialized the direction of user intents modeling and user satisfaction prediction in recommendation systems \cite{sheil2018predicting, guo2019buying, su2018user, mehrotra2019jointly}. In \cite{mehrotra2019jointly}, the authors identified different kinds of user intents in music recommendation system through interviews, surveys and quantitative approaches, and proposed a multi-level hierarchical model to predict user satisfaction based on their intents. In \cite{su2018user}, the authors classified user intents in product searching into three categories, including target product searching, decision making and new product exploration. They observed that different intents usually imply different interaction patterns, which can be further used to predict user satisfaction.

Advertiser intent modeling is different from these existing works in two main aspects. First, advertiser intent is more difficult to model than user intent. User intents in searching, recommendation system, and e-commerce platform are relatively explicit, accompanied with different behavior patterns, and can be identified into coarse discrete categories through surveys or interviews. According to our preliminary investigation, most of the advertisers' intents are a mixture of different optimization objectives, including PV, CTR, CVR, etc. Thus, instead of classifying advertiser intents into pre-defined discrete categories, we need to quantify advertiser intents in a unified continuous space. Furthermore, advertisers do not have appropriate channels to explicitly express their preferences over different performance indicators in most ad platforms. Second, advertiser's action is more complicated and dynamic than user behavior, the method to identify user intents is usually based on fine-grained interaction signals, such as mouse clicks and dwell time, according to the observation that user's behavior pattern is highly related to user's instant intent. However, advertiser's operations exhibit more complicated characteristics than user actions: different advertisers have different strategies for advertising, resulting in various action patterns;  action patterns of the same advertiser
could also change during an ad campaign due to the underlying evolving intents. Therefore, we may not only use fine-grained interaction signals to capture advertiser intent and satisfaction.
 
\subsection{User Click-through Rate Prediction}
In e-commerce platforms, click-through rate (CTR) indicates the user's interest and satisfaction over recommendation items. Recently, a number of works have been done in designing CTR prediction models via deep learning approaches \cite{covington2016youtube, cheng2016wide, zhai2016deepintent, shan2016deepcrossing, gai2017learning, qu2016pnn,zhou2018DIN,zhou2019DIEN}. Most of these models are based on the structure of embedding and multilayer perceptron \cite{zhou2018DIN}. To further improve the expression ability of the model, Wide \& Deep \cite{cheng2016wide} and DeepFM \cite{guo2017deepfm} used different modules to learn low-order and high-order features. PNN \cite{qu2016pnn} proposed a product layer to capture interactive patterns between inter-field categories. DIN \cite{zhou2018DIN} introduced the attention mechanism in CTR prediction problem to capture user's diverse interests. DIEN \cite{zhou2019DIEN} further improved DIN by designing an additional network layer to capture the sequential interest evolving process.

There are three main differences between advertiser satisfaction prediction and CTR prediction. First, we do not have a clear label of whether the advertiser is satisfied with the ad performance, while in CTR prediction, we can easily get the label by checking whether the user clicks the ads. Second, advertiser satisfaction is more complex than user click action. We not only need to consider advertisers' satisfaction but also explore their hidden intent, which is the underlying factors to determine advertiser satisfaction, while most CTR prediction models do not explicitly learn the user intent. In this work, we also identify advertiser intent during satisfaction prediction. Finally, CTR prediction models aim at building a correlation between a user and an item. However, advertiser comes to the ad platform to improve ad performance or product sale volume instead of a specific item. Thus we should focus on building a correlation between an advertiser and her ad performance.

\section{Preliminaries}
In this section, we first describe the interaction between advertisers and the displaying ad platform in Taobao. We then introduce the definitions of advertiser intent and satisfaction based on our observations from practical ad systems.

In Figure~\ref{fig:system}, we show an example of the interaction between advertisers and the Taobao displaying ad platform, in which advertisers can start or close an ad campaign at an appropriate time and take various actions during the ad campaign to achieve their marketing goals.
To initialize an ad campaign for an ad unit, the advertiser needs to choose a set of demographic tags and the corresponding basic bid prices to effectively and efficiently target the potential users. The advertiser would also select preferred ad positions (e.g., ad slots in ``Home Page'' or  ``After Shopping Page'') and a premium rate to improve the performance of ad exposure. The advertiser would like to declare a higher bid, i.e., $basic\ bid \times premium\ rate$ for the preferred ad positions. The advertisers participate in the ad auctions when targeting users visit the ad positions. If the ad unit wins the auction, the corresponding ad is shown to the users, and a certain price is charged to the advertiser once the user clicks the ad, following the cost per click (CPC) paradigm. Various auctions are conducted for the ad unit during the campaign, and the accumulated Key Performance Indicators (KPIs), such as CTR, Cost, ROI, and etc., are recorded in the report. After a certain period (e.g., one day or one week), the advertiser would check the KPIs report and conduct a series of actions to adjust the ad campaign when some performance indicators do not match her expectation. The potential operations of advertisers could be adding or deleting certain tags/positions or changing bid prices or premium rates. The ad unit with the adjusted features enters the ad auctions for competition and obtains new marketing performance.

Our goal in this work is to formally define the metrics to measure the advertiser marketing intent and satisfaction, build a connection between them via extracting the useful information from the above interactions, and identify advertiser intent and predict advertiser satisfaction based on the proposed metrics and connection.

\subsection{Advertiser Intent Modeling}
\label{sec:intent}
Advertisers use the Taobao ad platform with different marketing intents. However, existing work in the literature usually assumed the advertisers simply optimize a single objective, such as CTR or CVR, and the specific formulation of advertiser's practical and various intents is not well established. Our main focus in this subsection is to evaluate advertiser intent based on a survey and an intuitive observation from a mathematical optimization model.

To fully understand advertisers' marketing goals, we conduct a survey with questions about the major reasons of using Taobao displaying ad platform and the preferred performance indicators. We collect feedback from 791 advertisers, and show the results in Table~\ref{tab:advertiser_interview}.
\begin{table}[!t]
\caption{Summary of advertiser survey.}
\label{tab:advertiser_interview}
\centering
\begin{tabular}{cc|cc}
\hline
Marketing Intents & Ratio & Indicators & Ratio \\
\hline
Target potential users actively & 68\% & ROI, CVR & 86\%\\
Lever non-advertisement traffic & 52\% & CTR, PPC & 56\%\\
Cost-effective advertising & 46\% & CART, CLT & 55\% \\
\hline
\end{tabular}
\end{table}
We note that the ratio of each column in Table~\ref{tab:advertiser_interview} does not need to sum up to one, indicating the advertisers may have multiple intents and care for different performance indicators, such as ROI, CVR, CTR, CART (\emph{Add to Cart} number) and CLT (\emph{Collection} number), simultaneously. As a result, advertiser intent may not be simply characterized by a single indicator as in the existing work. Its formulation should reflect various concerns of the advertiser and contain the major indicators of the ad campaign.
We introduce an advertiser-specific weight vector $\mathbf{w} = [w_1, w_2, \cdots, w_n]^\top$ to define the advertiser's intent by assigning each indicator an importance weight, which represents the advertiser's preference over different indicators. One simple interpretation for the weight vector is that the advertiser may optimize her actions, such as adjusting bid or premium rate, through solving a multi-objective optimization problem. We can also derive the weight vector by solving the maximization problem for a specific indicator but with additional constraints for other indicators, which can be illustrated as follows:
\begin{equation}
\setlength{\abovedisplayskip}{1pt}
\setlength{\belowdisplayskip}{1pt}
\label{eqn:bid_simulator}
  \begin{gathered}
    \max_{bid}\ pv        \\
    s.t.\ click\geq \alpha,\ cost\leq \beta,\ ppc\leq \gamma, \\
  \end{gathered}
\end{equation}
where the performance indicators $pv, click, cost, ppc$ are functions of $bid$.
By employing Lagrange multipliers, we can derive the following Lagrangian function for the above optimization problem:

\begin{small}
\setlength{\abovedisplayskip}{1pt}
\setlength{\belowdisplayskip}{1pt}
\begin{flalign}
  \begin{aligned}\label{eqn:bid_sim_lagrange}
    \mathcal{L}(bid, \mathbf{\hat{w}})& = pv + w_{1}(click- \alpha ) + w_{2}(\beta - cost ) + w_{3}(\gamma - ppc) \\
    &=\mathbf{\hat{w}}^\top I + b,
  \end{aligned}
\end{flalign}
\end{small}
where $\mathbf{\hat{w}}=[1, w_{1}, -w_{2}, -w_{3}]^\top$, $I=[pv, click, cost, ppc]^\top$ and $b= - w_{1}\alpha + w_{2}\beta + w_{3}\gamma$.
The Lagrange multipliers have a nice interpretation in economics~\cite{ecnomic}. We can regard the Lagrange multipliers as some kind of prices for violating the indicator constraint. Therefore, we consider the Lagrangian function $\mathcal{L}(bid, \mathbf{\hat{w}})$ as the total benefit of declaring the $bid$, and the vector $[\mathbf{ \hat{w} }^\top, b]^\top$ as the weight vector $\mathbf{w}$ of advertiser's intent. However, there are several obstacles for optimizing~(\ref{eqn:bid_sim_lagrange}) explicitly. Due to information asymmetry between the ad platform and the advertiser: the ad platform has no direct knowledge of the constraints in~(\ref{eqn:bid_simulator}), e.g., the constraint parameters $\alpha, \beta, \gamma$. This may come from the scenario that the advertisers would not like to reveal this information due to privacy policy and business secret, or from the scenario that the advertisers may not have the feasible channels to express their optimization objectives in the current ad platform.

To resolve the lack of information about optimization parameters, we have a critical observation from the interaction between the ad platform and advertisers. The intents motivate advertisers to take different actions after checking the performance report, and thus their intents can be estimated by their historical action traces, such as deleting or adding tag actions, and the performance report. From the previous advertiser survey, we also observe the marketing intent of advertisers is usually consistent during a certain period. Thus, we can use the advertiser's historical information from a certain observation period to predict her current intent using our designed deep learning model in next section.

\subsection{Advertiser Satisfaction Modeling}
\label{sec:satisfaction}

It is quite challenging to determine whether the advertisers are satisfied with the performance of their ad units, compared with the scenario in the user's side, where we have direct and clear signals, such as click or buying actions, to infer their satisfaction over recommendation items. One potential metric to measure the advertiser satisfaction level is the Lagrange function in~(\ref{eqn:bid_sim_lagrange}), which is a weighted average of all advertising performance indicators. However, due to the above-mentioned information asymmetry, we are unable to calculate this metric exactly.

From empirical observation, we find that various indicators are highly related to advertiser satisfaction, which could be exploited to define an appropriate metric to infer the level of advertiser satisfaction via indirect signals. Churn label is a classical satisfaction metric to measure a customer's satisfaction over the subscribed services, e.g., various works~\cite{rosset2002customer, chamberlain2017customer, yang2018know, lu2019uncovering} define a churned customer in different contexts based on the following signals: (1) No placing an order over a year in ASOS platform \cite{chamberlain2017customer}, (2) No logging in over 30 consecutive days in TikTok platform \cite{lu2019uncovering}, (3) No activity over a week in Snapchat \cite{yang2018know}. Similarly, if an advertiser is unsatisfied with the performance of an ad unit, she would finally stop investing for the ad campaign. Thus we can adopt investment-related indicators to infer the advertiser's satisfaction to some extent, such as take rate (the ratio of cost to GMV) or simply the cost level. From the above discussion, we regard an advertiser is unsatisfied with the performance of an ad unit during a period $[l_0-l,l_0)$ if the cost in period $[l_0,l_0+l)$ is less than or equal to $\epsilon$, which is a small amount of money close to 0. Otherwise, the advertiser is satisfied with (or at least not so unsatisfied) with the ad unit. Usually, we do not set $\epsilon$ equal to $0$ because an advertiser may leave before her budget runs out. Due to the performance delay in the ad platform, we use the cost level in a later period to evaluate the satisfaction level in a previous period. With this definition, we can have a direct and easy-calculated signal to obtain the satisfaction label for the data.

\section{Deep Satisfaction Prediction Network}
\begin{figure*}[t]
  \centering
  \centerline{\includegraphics[width=1.90\columnwidth]{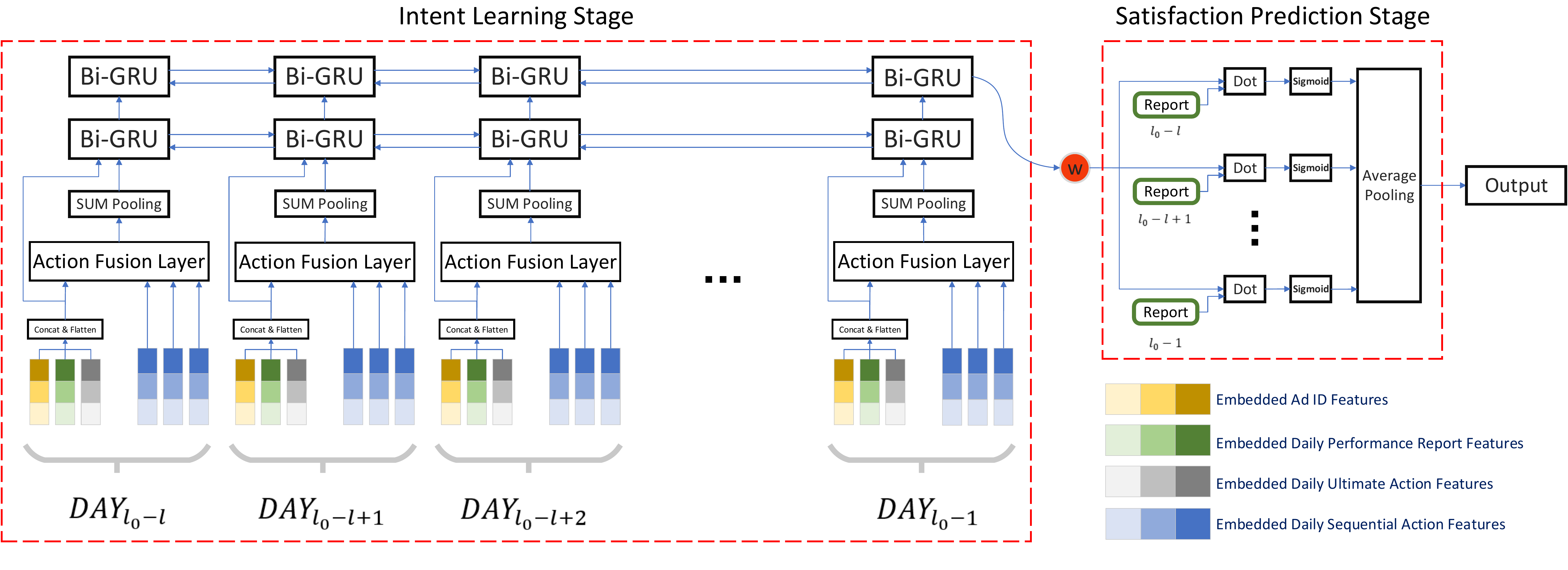}}
  \caption{Deep Satisfaction Prediction Network. The left part is the intent generation stage, which computes the intent vector $\mathbf{w}$, the right part is the satisfaction prediction stage, which predicts advertiser satisfaction based on ad unit's performance report and advertiser's intent vector.}
  \label{fig:DPSN}
\end{figure*}

In this section, we propose a novel deep network architecture, Deep Satisfaction Prediction Network (DSPN), for the advertiser intent learning and satisfaction prediction. As shown in Figure~\ref{fig:DPSN}, DSPN is a supervised learning framework with two stages: intent learning stage and satisfaction prediction stage. The learnable parameters are all deployed in the intent learning stage, which extracts information from the advertiser's historical data to learn intent vector $\mathbf{w}$, which is forwarded to the second stage. The satisfaction prediction stage combines the intent vector $\mathbf{w}$ and advertiser's historical performance report to calculate whether an advertiser is satisfied over the performance of ad units.

\subsection{Input Features}
We first investigate which types of features related to advertiser's intent and satisfaction. As heterogeneous advertisers may have various intents and satisfaction levels for different ad units, the prediction model needs to involve the ID features of advertisers and ad units. It is obvious that the specific values of KPIs in the report are highly related to the advertiser's satisfaction levels. Thus, we also need to consider the performance indicators during the training of prediction model. Advertisers' behavior patterns have certain relation with their satisfaction over the ad performance. For example, when we count the number of actions of an advertiser for a specific ad unit during an observation period with length $10$, we find that the average number of modification actions in positive label (satisfied) is $11.179$, while the mean in negative label (unsatisfied) is $9.790$, indicating that satisfied advertisers tend to achieve the desired ad performance through frequent adjustments of bid prices and premium rates. Moreover, for the number of ``add tag'' actions, the mean of positive label is $1.274$, and negative label $1.267$, which are close. However, for the number of ``delete tag'' actions, the mean of positive label is 0.178 more than that of negative label, which is 0.109. These findings indicate that satisfied advertiser tends to adjust her strategy by deleting demographic tags which she thinks not so important after some trials. Therefore, we also need to consider the actions of advertisers in the model.

From the above discussion, the input of DSPN contain three types of features: ID features, performance indicators and action features, from an observation period. We further divide action features into ultimate actions and sequential actions. The ultimate actions represent the advertiser's strategy at the end of the day, containing tags' final bid prices and ad positions' premium rates. The sequential actions represent the intermediate actions the advertiser has done during the day to achieve her marketing intents. The specific formats of the data sets are described in Section~\ref{sec:exp}.

\subsection{Embedding Layer}

We transform our sparse features into low-dimensional dense features through embedding. For categorical feature, e.g. ID features, we convert them into a dense real-valued embedding vector $\mathbf{m}$. For features with both categorical and numerical information, we present them with $\mathbf{e} = v \odot \mathbf{m}$, where the embedding vector $\mathbf{m}$ indicates categorical information, the value $v$ is the normalized value indicating numerical information, and $\odot$ is the element-wise product operator. Furthermore, the value $v$ of performance indicator feature is the value of the corresponding KPI; the value $v$ of ultimate action feature is the value of bid price/premium rate at the end of the day; the value $v$ of sequential action feature is the value of bid price/premium rate after the action minus the value before the action.

After the embedding layer, features of the same type like ID features are concatenated into a vector. Fully-connected layers are used for reshape the vector when necessary. For sequential action features, we further divide them into two groups: tag features and ad position features. Each of the group contains a sequence of action features within the day, and we connect the embedding vectors of these actions chronologically to form a list represented by $E=[\mathbf{e}_1, \mathbf{e}_2,...,\mathbf{e}_{n_a}] \in \mathbb{R}^{n_e \times n_a}$, where $\mathbf{e}_i$ is the embedding vector of the $i$-th sequential action with dimension $n_e$, and $n_a$ represents the number of actions in this group within a day. The embedding layer is trained at the same time with the model during the training process. We use sum/average pooling to transform multiple embedding vectors into a fixed-length vector when necessary.

\subsection{Intent Learning Stage}
In the intent learning stage, we employ multilayer bidirectional RNNs on the top to learn the advertiser intent based on information from $l$ sequential days. We use a fusion layer to capture important pattern in sequential actions for each day in the bottom.

\textbf{Action Fusion Layer.} Advertiser's sequential actions such as \emph{``add a tag with price x''} or \emph{``delete a tag''} contain rich information of her strategies or preferences over various KPIs. We use a fusion layer to get a summary of advertiser's sequential actions within each day. Different approaches like sum pooling, average pooling or attention mechanism \cite{bahdanau2014neural} can be used in the fusion layer. Since advertisers may try different actions before they finally figure out the correct actions to do within a day, we suggest employ the attention mechanism to identify important actions within the action sequence, considering the advertiser's current situation. In detail, the attention mechanism is formulated as:
\begin{equation}
\setlength{\abovedisplayskip}{1pt}
\setlength{\belowdisplayskip}{1pt}
    \overline{V} = \text{softmax}(\frac{V Q^\top}{\sqrt{n_a}}) \cdot V
\end{equation}
where $V \in \mathbb{R}^{n_a \times n_e}$ is the sequential action matrix of one day and $V=E^\top$, values of $n_a$ and $n_e$ represent the number of actions in each day and the embedding dimension, respectively. The matrix $Q \in \mathbb{R}^{n_a \times n_e}$ is the summary of the advertiser's ID features, daily performance report and daily ultimate action information on the same day, and is reshaped to the same size as $V$. One can obtain $Q$ through MLP, RNN or other structures. With the attention mechanism, we can learn an advertiser-specific representation for the action pattern. Matrix $\overline{V}$ is given to the next level together with other feature representations after pooling.

\textbf{RNN Layer.} Recurrent neural network models sequential data explicitly~\cite{lipton2015critical} and gives us inspiration to exploit the advertiser intent based on her historical information. Imagine the procedure that an advertiser reviews her recent ad performance and actions, and adjust the actions to obtain updated ad performance if her marketing intent is not well fulfilled. Thus, there are some underlying relation between advertiser actions and advertiser intent. We employ RNN to investigate the dynamic representation of advertiser's daily information related to advertiser intent. To balance efficiency and performance, we employ GRU~\cite{chung2014GRU} as GRU overcomes vanishing gradient problem and is faster than LSTM~\cite{hochreiter1997LSTM}. The formulations of GRU are listed as follows:
\begin{align}
\setlength{\abovedisplayskip}{1pt}
\setlength{\belowdisplayskip}{1pt}
& r_t=\sigma(W_r e_t + U_r h_{t-1} + b_r)\\
& z_t=\sigma(W_z e_t + U_z h_{t-1} + b_z)\\
& h_t=\text{tanh}(W_h e_t + U_h (r_t \odot s_{t-1}) + b_h)\\
& s_t=z_t \odot s_{t-1} + (1-z_t) \odot h_t
\end{align}
where $e_t$ is the representation of $t$-th day's report and action information from the previous level, $s_t$ is the $t$-th hidden states, $\sigma$ is the sigmoid function and $\odot$ is the element-wise product operator. Because advertisers may review the historical information back and forth, we would like the representation of each day to summarize not only the preceding days, but also the following days. Hence, we propose to use two layers of bidirectional recurrent network (Bi-GRU) \cite{schuster1997birnn}. For each Bi-GRU layer, the hidden state of the $t$-th day can be represented by concatenating the hidden state in the forward GRU and the backward GRU as $h_t=[\overrightarrow{h}_t ,\overleftarrow{h}_t]$. We obtain the representation of $\mathbf{w}$ by adding the $l_0-1$-th day's hidden state in the second Bi-GRU layer as $\mathbf{w}=\overrightarrow{h}_{l_0-1} + \overleftarrow{h}_{l_0-1}$.

\subsection{Satisfaction Prediction Stage}
In satisfaction prediction stage, we build a connection between advertiser intent vector and advertiser satisfaction level. According to Section~\ref{sec:intent}, there exists a positive correlation between $\mathbf{w}^\top  I_i$ and advertiser satisfaction, where $I_i$ represents the performance report in day $i$. Following this intuition, we assign the probability of the advertiser being satisfied with the ad performance during the period $[l_0-l, l_0)$ as:
\begin{equation}
\setlength{\abovedisplayskip}{1pt}
\setlength{\belowdisplayskip}{1pt}
    p(x)=\frac{1}{l}\sum_{i=l_0-l}^{l_0-1} \sigma(\mathbf{w}^\top I_i),
\end{equation}
where $\sigma$ is sigmoid function and $\mathbf{w}$ is the advertiser intent vector learned from the previous stage. The loss function of the prediction model is defined as follows:
\begin{equation}
\setlength{\abovedisplayskip}{1pt}
\setlength{\belowdisplayskip}{1pt}
L=-\frac{1}{N}\sum_{(x,y) \in \mathcal{D}}^{N} (y log p(x) + (1 - y) log (1 - p(x)),
\end{equation}
where $\mathcal{D}$ is the training set with size $N$, $x$ is the input of the model and $y$ is the label.

\section{Experiments}
\label{sec:exp}

In this section, we evaluate DSPN in detail and summarize our experiments as follows: (1) We verify the effectiveness of DSPN in intent identification and satisfaction prediction, and discuss the impact of different data features through offline experiments. (2) Analyses of intent vector $\mathbf{w}$ show that $\mathbf{w}$ can fully capture the advertiser intent in practice. (3) Online evaluation shows that our model performs well for different advertiser satisfaction modeling tasks and has stable performance with around $0.9334 \pm 0.0028$ AUC counting in $30$ consecutive days.

\subsection{Experiment Setup}

The dataset we used is collected from Alibaba displaying ad system. There are about $1$ million ad units from around $1.6 \times 10^5$ advertisers in our dataset, and each ad unit forms a unique sample. The ad units we selected are those with cost larger than $10$ units in period $[l_0-l,l_0)$. We randomly divide the data set into a training set and a test set according to a $9: 1$ ratio. Each data sample contains an ad unit's information from the observation period $[l_0-l,l_0)$ with length $l=10$, and the satisfaction label is marked for each data sample using the definition in Section~\ref{sec:satisfaction} with the parameters $l=10$ and $\epsilon=10$. We set these parameters based on our business experience and data analysis. Specifically, when $l$ is too small, it may not reflect real satisfaction phenomenon of users; when $l$ is too large, it would involve many already churned ad units during the observation period, or the downstream tasks would wait a long time (at least $l$ days later) to get the satisfaction predicted results. The parameter $\epsilon$ is a small amount of money that can prevent noisy data introduced by ad performance delay, and enable the model to do early churn warning before budget runs out. The details of main features in the dataset are summarized as follows:

\textbf{ID Features.} ID features contain basic profile and identity information about ad unit, product category and advertiser, and are used to provide personalized information in model training. 

\textbf{Report Features.} Report features refer to the performance indicators of an ad unit within each day from the observation period. There are $15$ indicators of report features in total, including Cost, CTR, CVR, ROI, and etc.

\textbf{Action Features.} Action features contain the ultimate action feature and sequential action feature of the advertiser within the observation period. The detailed data format for action features are described in Table \ref{tab:davn_input3}.

We train DSPN with batch-GD algorithm with batch size $32$, using the Adam optimizer \cite{kingma2014adam}. The dimension of the hidden state in the first Bi-GRU layer is $18$, the dimension of the hidden state in the second Bi-GRU layer is $16$, which is the same as the dimension of $\mathbf{w}$. Different categorical features have different embedding dimensions in DSPN from $3$ to $18$, which are fine-tuned to achieve good results. The source code is available online\footnotemark.

\footnotetext{https://github.com/liyiguo95/DSPN}

\subsection{Comparative Models}
The problem of advertiser satisfaction prediction we discussed is new in both academia and industry communities, and to the best of our knowledge, there are not related models in the literature. Considering that the prediction models for user-side problems are widespread and mature, we carefully select some widely used prediction models in user click-through rate prediction, slightly revised them to adopt to the scenario of advertiser satisfaction, and regard them as our baselines.

\textbf{Embedding \& MLP Model:} The Embedding \& MLP model is a basic model of most subsequently developed deep networks \cite{covington2016youtube,cheng2016wide,qu2016pnn,guo2017deepfm,zhou2018DIN} for CTR modeling. It contains three parts: embedding layer, pooling and concat layer, multilayer perceptron.

\textbf{Wide \& Deep \cite{cheng2016wide}:} Wide \& Deep model has been proved to be effective in recommendation systems. It consists of two parts: (1) Wide model, which is a linear model handles the manually designed features. (2) Deep model, which employs the Embedding \& MLP Model to learn nonlinear relations among features automatically.

\textbf{PNN \cite{qu2016pnn}:} PNN can be viewed as an improved version of the Embedding \& MLP Model by explicitly introducing a product layer after embedding layer to capture high-order feature interactions.

\textbf{DeepFM \cite{guo2017deepfm}:} DeepFM imposes factorization machines as the ``wide'' part and multilayer perceptron as the ``deep'' part. The two parts in DeepFM are trained together and share the input.

\textbf{DIN \cite{zhou2018DIN}:} 
Based on the Embedding \& MLP Model, DIN uses the attention mechanism to activate related user behaviors. In our experiments, we implements DIN to learn the inner relationship between the advertiser profiling and daily features.

\begin{table*}[h]
\renewcommand{\arraystretch}{0.8}
  \caption{Action features of Alibaba dataset.}
  \label{tab:davn_input3}
  \begin{tabular}{l|c|c|l}
    \toprule
    Feature Type&Target Type&Data Format&Data Explanation\\
    \midrule
    \multirow{2}{*}{Ultimate Action}&Tag&$ (tag\_type, bid\_price)$& bid for tags end of the day \\
     \cline{2-4}
     &Ad Position& $(position\_type, premium\_rate)$& premium rate for ad positions end of the day\\
    \cline{1-4}
    \multirow{6}{*}{Sequential Action}&\multirow{3}{*}{Tag} &$(tag\_type, price,time)$ & add a tag with a bid\\ 
     \cline{3-4}
     && $ (tag\_type, old\_price, new\_price,time)$ &change bid for a tag\\
     \cline{3-4}
     && $ (tag\_type, current\_price,time)$&delete a chosen tag\\
    \cline{2-4}
    \cline{2-4}
   &\multirow{3}{*}{Ad Position} &$(position\_type, rate, time)$&add an ad position with a premium rate \\ 
     \cline{3-4}
     && $(position\_type, old\_rate, new\_rate, time)$&change premium rate for an ad position\\
    \cline{3-4} 
     && $(position\_type, current\_rate, time)$&delete a chosen ad position\\
  \bottomrule
\end{tabular}
\end{table*}

\begin{table}
\setlength{\abovedisplayskip}{1pt}
\setlength{\belowdisplayskip}{1pt}
\renewcommand{\arraystretch}{0.6}
\caption{Comparisons of different models.}
\label{tab:experiment_result}
\begin{tabular}{lcccc}
\toprule
Model & AUC &  ACC \\
\midrule
MLP & 0.8501  & 0.7975 \\
Wide \& Deep & 0.8507 & 0.7968\\
PNN & 0.8519  & 0.7989\\
DeepFM & 0.8538 & 0.8005\\
DIN & 0.8562 &  0.7951 \\
DSPN without GRU & 0.9001  & 0.8414 \\
DSPN with multilayer GRU & 0.9415  & 0.8917 \\
DSPN with multilayer bidirectional GRU & \textbf{0.9437}  & \textbf{0.8928} \\
\bottomrule
\end{tabular}
\end{table}
\subsection{Experimental Results}
\textbf{Comparative Results.} Since the above mentioned models have different structures, to tune each baseline model, we use sum pooling, average pooling, and concatenation to integrate different features according to the specific model structure. The metrics used are Area Under the Curve (AUC) and Accuracy (ACC). From Table~\ref{tab:experiment_result}, we observe that DSPN outperforms other models in the problem of advertiser satisfaction prediction, achieving AUC of $0.9437$ and ACC of $0.8928$. We verify the effectiveness of DSPN through ablation studies of model structure. As shown in Table~\ref{tab:experiment_result}. DSPN with multilayer bidirectional GRU layers is superior to DSPN without GRU and DSPN with multilayer GRU. This result demonstrates that DSPN with bidirectional GRU layers in the first stage is more effective to extract useful patterns from the historical information than other designs. For the attention mechanism in the action fusion layer, we also test sum pooling, average pooling as the alternative fusion layer, and the results are similar. To shed some light on the advantages of the attention mechanism in broad scenarios: (1) We delete other features and only use sequential actions to perform the control experiments. (2) We also deploy DSPN in a similar task related to advertiser's satisfaction: predicting whether an ad unit's cost will increase in the next week, which highly relies on sequential action information. Compared with other strategies in the fusion layer, attention mechanism achieves better performance in both tasks with AUC increasing more than $3.0\%$ and $1.6\%$ respectively. We conclude that attention mechanism makes DSPN more robust to the tasks with relation to sequential action information.

\textbf{Impact of Data Features.} To better understand the roles that different data features play in advertiser satisfaction prediction, we conduct two types of tasks using DSPN. The first task is to predict the advertiser satisfaction without a selected data feature, while the second task only uses the selected data feature for prediction. We show the results in Table~\ref{tab:ablation_study_feature}, and can see that each data feature has a positive contribution to the performance of DSPN. ID feature contributes the least since it only provides basic information of an entity. However, we still maintain ID features so that information can be shared among the same entity. For example, ad units belong to certain categories may have good sale in some special time or areas. Thus, these ad units can then share the performance indicators for satisfaction prediction. Both daily report feature and action feature play important roles in satisfaction prediction. This observation shows that the advertiser satisfaction mainly depends on daily ad performance, which proves the necessity of optimizing the ad results under the guideline of advertiser's intent.

\subsection{Intent Results Analysis}
In this subsection, we show the validity and the effectiveness of the weight vector $\mathbf{w}$ learned from the first stage in representing various advertiser intents. We also show the necessity of considering advertiser intents when predicting their satisfaction.

\textbf{Effectiveness of weight vector $\mathbf{w}$ in profiling intents.} 
We perform a \emph{K-Means} clustering analysis on the weight vector $\mathbf{w}$ in the test set and investigate whether $\mathbf{w}$ can capture the advertisers' various intents. We use \emph{Euclidean distance} as the cluster metric in \emph{K-Means} clustering and use \emph{PCA} to reduce the dimensions of $\mathbf{w}$ from $16$ to $2$, then we visualize the sample distribution of different clusters in Figure~\ref{fig:cluster}. We find the obtained cluster centers well match the typical advertisers in practical ad systems. 

\textbf{Cluster 1: Active ad units.} Advertisers in this cluster are high quality customers to the ad platform. Most of advertisers feel satisfied with the ad units in this cluster, because the ad units have a stable and good performance in overall indicators.

\textbf{Cluster 2: Maximizing Impression.} Advertisers in this cluster care more about PV and Click number, and they also care about cost. This indicates that advertisers would like to maximize impression or Click number in a cost-effective way.

\textbf{Cluster 3: Maximizing Revenue.} Advertisers in this cluster are insensitive to the cost, and consider more on ROI and CVR. These ad units usually target the loyal customers or members in the advertiser's Taobao shop.

\textbf{Cluster 4: Tail ad units.} Advertisers in this cluster usually do not spend too much money in total. Performance indicators like click and pay amount have a positive impact in their satisfaction, while cost has a negative impact in their satisfaction.

\begin{table}
\setlength{\abovedisplayskip}{1pt}
\setlength{\belowdisplayskip}{1pt}
\renewcommand{\arraystretch}{0.6}
  \caption{Impact of different features. Results of tasks without a selected feature are on the left, and results of tasks only with the selected feature are on the right.}
  \label{tab:ablation_study_feature}
\begin{tabular}{lcc|lcc}
    \toprule
    Task & AUC & ACC & Task & AUC & ACC\\
    \midrule
    w/o ID & 0.9428 & 0.8924 & w ID & 0.8241 & 0.7669\\
    w/o Action & 0.9410 & 0.8903 & w Action & 0.9300 & 0.8853 \\
    w/o Report & \textbf{0.9336} & \textbf{0.8860} & w Report & \textbf{0.9368} & \textbf{0.8890}\\
\bottomrule
\end{tabular}
\end{table}

\begin{table}
\renewcommand{\arraystretch}{0.6}
  \caption{Results of intent effectiveness.}
  \label{tab:cluster}
  \begin{tabular}{lccc}
    \toprule
     &ACC In-cluster &ACC Other-clusters& Equal Ratio\\
    \midrule
    Cluster 1&0.9962& 0.6286 & 0.9578 \\
    Cluster 2&0.7745&0.7071 &0.8152 \\
    Cluster 3&0.9657&0.6384 &0.8048 \\
    Cluster 4&0.8702& 0.1850&0.5070 \\
    All & 0.8734&0.5669 & 0.7712 \\
  \bottomrule
\end{tabular}
\end{table}

\begin{figure}
\setlength{\abovedisplayskip}{1pt}
\setlength{\belowdisplayskip}{1pt}
\centering
    \begin{subfigure}{0.25\textwidth}
    \centering
        \includegraphics[width=1.0\linewidth]{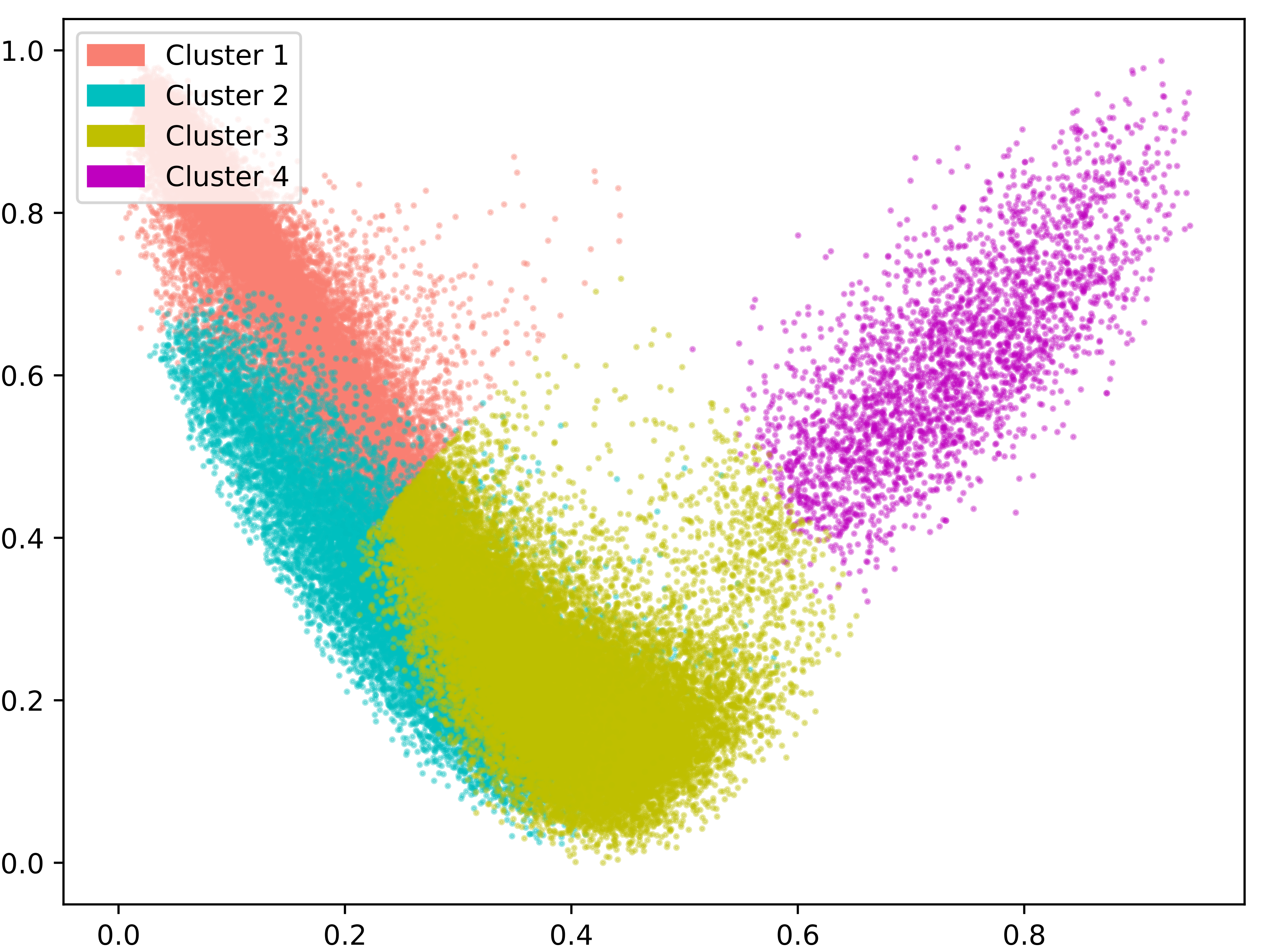}
        \caption{Positive Samples.}
        \label{fig:cluster_neg}
    \end{subfigure}%
    \begin{subfigure}{0.25\textwidth}
    \centering
        \includegraphics[width=1.0\linewidth]{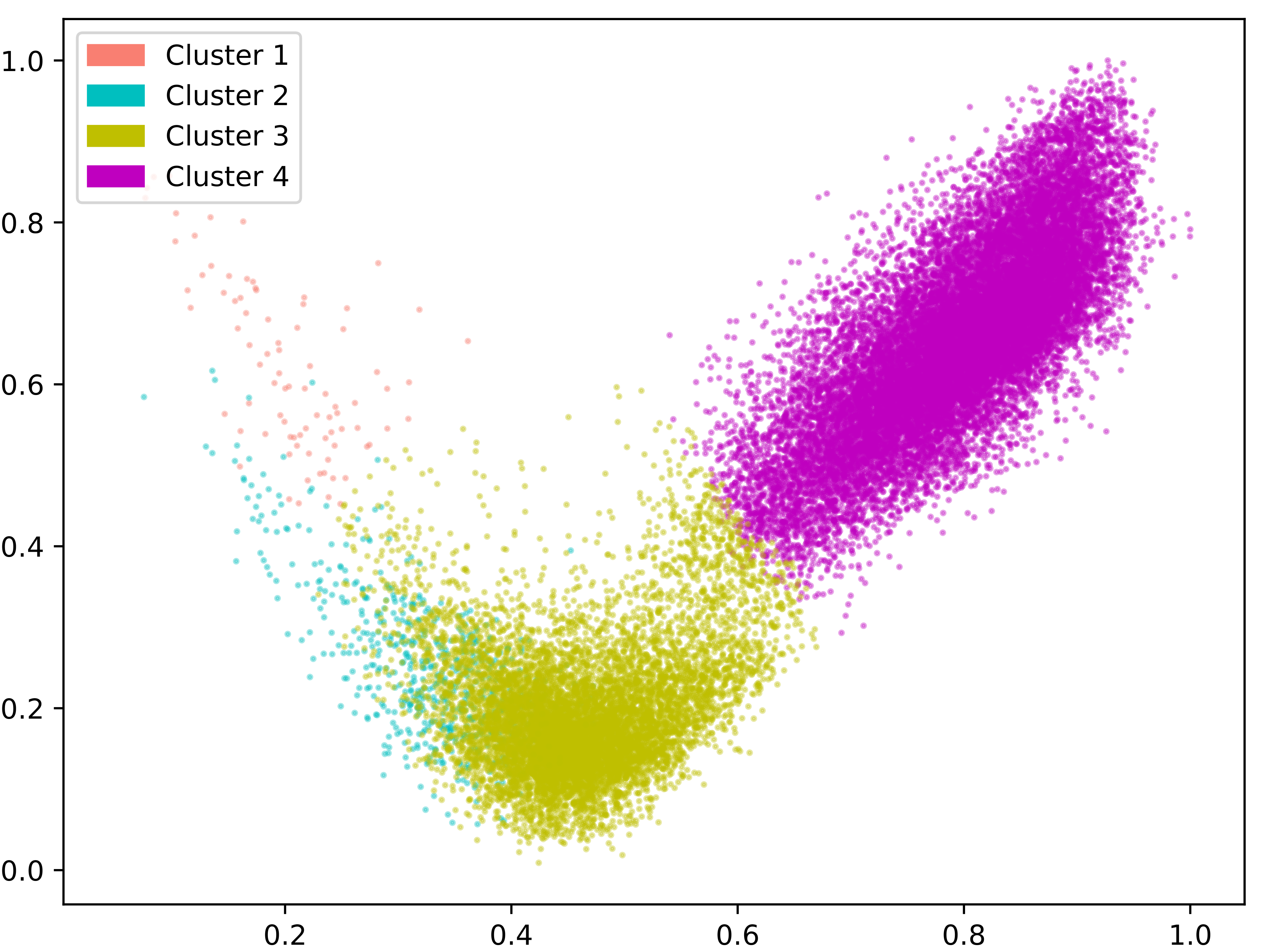}
        \caption{Negative Samples.}
        \label{fig:cluster_pos}
    \end{subfigure}
\caption{Distribution visualization of intent vector $\mathbf{w}$. We use \emph{PCA} to reduce the dimensions of $\mathbf{w}$ from $16$ to $2$, and visualize the sample distribution of different clusters after normalization.}
\label{fig:cluster}
\end{figure}

\textbf{Importance of advertiser intent in satisfaction prediction.} We design three experiments to figure out how different intents can significantly influence the performance of advertiser satisfaction, and show the results in Table~\ref{tab:cluster}. The first column in Table~\ref{tab:cluster} shows the accuracy of predicting each advertiser's satisfaction label with her report data $I$ and $\mathbf{w}$ from the center of the cluster she belongs to. The results show that these cluster centers can well predict the satisfaction of advertisers belonging to the same cluster, i.e., the advertisers having similar marketing intents. In the second experiment, we use each cluster center to predict the satisfaction label of advertisers in other clusters. The results in the second column of Table~\ref{tab:cluster} show that the prediction accuracy is far worse than that in the same cluster as we mentioned above, indicating advertiser intents are different among clusters. In the third experiment, we randomly select $5000$ samples in each cluster. For each sample, we find its closest sample in other clusters in terms of \emph{Euclidean distance}, and we compare the labels of these two samples to find out whether advertisers with similar reports have the same satisfaction label. The results in the last column of Table~\ref{tab:cluster} prove that even with similar advertising reports, advertisers' satisfaction can be different due to the differences in intents. The only exception is the samples similar to Cluster $1$, which has accuracy of $95.78\%$. This is because advertisers usually feel satisfied with ad units with good overall performance.

The intent vector can be used as auxiliary information for downstream tasks. For example, we could use $w_1 pCTR + w_2 pCVR + w_3 pCost$ as the ranking index in tag recommendation, where $pCTR$, $pCVR$ and $pCost$ are the predicted indicators and $w_i$ comes from the intent vector learned from advertisers. Weight average over indicators can also be used as the objective of real-time bidding algorithms.

\subsection{Online Evaluation}

Our model can be deployed in different business scenarios related to advertiser satisfaction, such as whether an advertiser will be churned or not, whether the cost or take rate of an ad unit or advertiser will increase in the following week, or the advertiser's attitude towards an adjust performance report. We have currently deployed our model as a key component in our online advertiser churn prediction system to serve different downstream optimization tasks, such as coupon distribution for unsatisfied advertisers, for more than four months. We use DSPN to predict the advertiser's satisfaction of an advertiser over their all ad campaigns, which is a slight different from our offline setting, in which we use DSPN to predict the satisfaction over an ad unit. To overcome such a difference, we use the features like performance report and action information of an advertiser instead of an ad unit, to train DSPN model. The average AUC of DSPN for the advertiser churn prediction task in one month is around $0.9334 \pm 0.0028$. 

In online evaluation, we would get the actual churn result $10$ days later based on the churn label, thus the training set and test set have a time gap of at least $10$ days. To shed some light on the model's performance at the moment, we randomly divide the data set into a training set and a validation set according to a $9: 1$ ratio. The average AUC in the validation set of the same $30$ days is $0.9350 \pm 0.0033$, which is almost the same as the actual average AUC ($0.9334 \pm 0.0028$). This result indicates that the model can well handle the performance loss caused by the time gap. In particular, our model has precision of $68.91\%$ and recall of $51.96\%$ in predicting the loyal advertisers who have consumption in the recent $10$ days but will not have consumption in the next $10$ days, which largely outperforms existing system's main strategy, a rule-based method. In our nearest optimization through optimizing ad performance, the logging and recharging rate of $8500$ advertisers increase by $0.7\%$ and $1.0\%$ respectively than $8500$ homogeneous advertisers without optimization.

\section{Conclusion}
In this paper, we have investigated the problem of advertiser intent learning and satisfaction prediction. Based on advertiser survey, empirical observations and mathematical analyses, we used a weight vector over advertising performance indicators to model the advertiser intent. Considering that the satisfied advertisers would continue to invest in ad campaigns, we proposed a metric related to the change of cost to evaluate the advertiser satisfaction. We then designed a deep learning network, namely DSPN, to identify advertiser intent and predict advertiser satisfaction, using the profile information of advertisers and ad units, performance indicators and sequential actions. Experimental results on an Alibaba advertisement dataset have shown the superiority of DSPN compared with the baseline models in terms of AUC and accuracy, and the effectiveness of using the weight vector to interpret the advertiser's intent. DSPN has been deployed in the online advertiser churn prediction system in Alibaba, and has helped to avoid the unsatisfied advertisers to leave the ad platform.

\bibliographystyle{ACM-Reference-Format}
\bibliography{CIKM2020_DSPN_Guo}

\end{document}